\documentclass[rc]{jjap3}
\usepackage{txfonts}

\title{First-principles study on magnetic tunneling junctions with semiconducting CuInSe${}_2$ and CuGaSe${}_2$ barriers}
\author{Keisuke Masuda$^{1}$ and Yoshio Miura$^{1,2,3,4}$\thanks{E-mail: MIURA.Yoshio@nims.go.jp}}
\inst{$^{1}$Research Center for Magnetic and Spintronic Materials, National Institute for Materials Science (NIMS), Tsukuba, Ibaraki 305-0047, Japan \\
$^{2}$Kyoto Institute of Technology, Electrical Engineering and Electronics, Kyoto 606-8585, Japan \\
$^{3}$Center for Materials Research by Information Integration, National Institute for Materials Science (NIMS), Tsukuba, Ibaraki 305-0047, Japan \\
$^{4}$Center for Spintronics Research Network (CSRN), Graduate School of Engineering Science, Osaka University, Toyonaka, Osaka 560-8531, Japan }
\abst{We theoretically investigate two magnetic tunneling junctions (MTJs) with different semiconductor barriers, CuInSe${}_2$ (CIS) and CuGaSe${}_2$ (CGS), sandwiched between Fe electrodes. We find that $\Delta_1$ wave functions provide dominant contributions to spin-dependent tunneling transport in both CIS- and CGS-based MTJs. We also find that the CGS-based MTJ has a much higher magnetoresistive (MR) ratio than the CIS-based MTJ, which indicates that a higher MR ratio is expected for a higher Ga concentration $x$ in the recently reported CuIn${}_{1-x}$Ga${}_x$Se${}_2$-based MTJs. Furthermore, we show that the CIS- and CGS-based MTJs have much smaller resistance-area products ($RA$) than the conventional MgO-based MTJs.}

\begin{document}
\maketitle

Magnetoresistive (MR) devices with high MR ratios and small resistance-area products ($RA$) are required for realizing read sensors of ultrahigh-density hard disk drives and Gbit-class spin transfer torque magnetoresistive random access memories (STT-MRAMs). Various attempts have been made to reduce the $RA$ of MgO-based magnetic tunneling junctions (MTJs)\cite{2004Parkin-NatMat,2004Yuasa-NatMat} to less than $1\, \Omega\, \mu {\rm m}^2$ while keeping high tunneling magnetoresistance (TMR) ratios. Elaborate techniques to deposit ultrathin MgO barriers ($\sim$$1\,{\rm nm}$) have been established,\cite{2007Choi-APL,2008Isogami-APL,2011Maehara-APEX} enabling the reduction in $RA$ to $\sim$$1\, \Omega\, \mu{\rm m}^2$ while keeping the high MR ratio of around $200\%$ at room temperature. On the other hand, the use of half-metallic Co-based Heusler alloys as ferromagnetic (FM) electrodes increased the MR ratio in small-$RA$ current-perpendicular-to-plane giant magnetoresistive (CPP-GMR) devices.\cite{2006Yakushiji-APL,2008Nikolaev-JAP,2008Furubayashi-APL,2011Sato-APEX,2013Li-APL,2016Jung-APL} The highest MR ratio reported so far is $82\%$ at room temperature for the ${\rm Co_{2}FeGa_{0.5}Ge_{0.5}/Ag/Co_{2}FeGa_{0.5}Ge_{0.5}}$ system with $RA$ of $\sim$$40\, {\rm m}\Omega\, \mu{\rm m}^2$.\cite{2016Jung-APL} However, this $RA$ value is too small to obtain sufficiently high voltage output under the current for read sensor applications. Most recently, Kasai {\it et al}. reported high MR ratios of $40\%$ at room temperature and $100\%$ at $8\,{\rm K}$ in the MTJ using the compound semiconductor ${\rm CuIn_{0.8}Ga_{0.2}Se_2}$ (CIGS) with the chalcopyrite crystal structure as a barrier in combination with ${\rm Co_{2}FeGa_{0.5}Ge_{0.5}}$ FM layers.\cite{2016Kasai-APL} This is the first observation of high MR output for MTJs with a compound semiconductor barrier. Since the band gap of the CIGS is much smaller than that of the insulator MgO, a smaller $RA$ is expected. Actually, sufficiently small $RA$ values ranging from $0.3$ to $3\, \Omega\, {\mu {\rm m}}^2$ were observed in the CIGS-based MTJs. Moreover, these MTJs are expected to have high controllability and high breakdown voltage because their barrier thicknesses ($\sim$$2\,{\rm nm}$) are two times larger than those of the above-mentioned MgO-based MTJs with comparably small $RA$. Such semiconductor barriers open up another path towards realizing both small $RA$ and high MR output.

From the theoretical point of view, several previous studies have focused on the transport properties of the MTJs with semiconductor barriers. MacLaren {\it et al.}\cite{1999MacLaren-PRB} studied MTJs consisting of Fe electrodes and a ZnSe semiconductor barrier within the first-principles calculations using the layer Korringa$-$Kohn$-$Rostoker approach. They showed that the system has spin-dependent tunneling transport properties, in which $\Delta_1$ bands provide dominant contributions. From their data, the MR ratio was estimated to be $\sim$$500 \%$ for the barrier of 50 atomic units ($\sim$$2.6\, {\rm nm}$). In another theoretical work, Aut\`es {\it et al.}\cite{2010Autes-PRB} discussed the MTJ composed of a GaAs barrier sandwiched between Fe electrodes. They calculated spin-dependent conductance and predicted a maximum MR ratio of nearly $400\%$ for around 10 atomic layers of GaAs. Moreover, they considered the spin-orbit interaction and found that the effect of such interaction is significant for sufficiently thick barriers ($\gg$$20$ atomic layers, $\sim$$2.8\,{\rm nm}$). Although these theoretical approaches predicted high MR ratios, such a notable output has not been experimentally observed in the ZnSe- and GaAs-based MTJs; only a low MR ratio ($<$$2\%$) has been reported in the GaAs-based MTJs.\cite{2002Kreuzer-APL,2006Moser-APL} On the other hand, no theoretical studies have focused on CIGS-based MTJs in spite of the recent report of high MR output. To understand the origin of such a high MR ratio in small-$RA$ TMR devices, a theoretical understanding of the transport properties of FM/CIGS/FM MTJs is essential.

In this work, we study transport properties of two MTJs with different barriers, ${\rm CuInSe}_2$ (CIS) and ${\rm CuGaSe}_2$ (CGS), which are the terminal compounds of a ${\rm CuIn}_{1-x}{\rm Ga}_{x}{\rm Se}_{2}$ mixed crystal. Since the band gap of ${\rm CuIn}_{1-x}{\rm Ga}_{x}{\rm Se}_{2}$ continuously increases as $x$ increases, CIGS is located between CIS and CGS not only chemically but also physically. Therefore, the present study of the CIS and CGS terminal compounds is adequate for obtaining sufficient information on the CIGS-based MTJ. As electrodes, we adopt ferromagnetic bcc Fe with a well-known band structure. Since the $a$-axis length of CIS (CGS) is almost twice as large as that of bcc Fe, the lattice mismatch between them is expected to be small. Note that we do not consider the effect of the spin-orbit interaction in this work, because we focus on a thin barrier of $\sim$$2\, {\rm nm}$, in which the effect of the spin-orbit interaction is sufficiently small as shown in a related study on an MTJ with a GaAs barrier.\cite{2010Autes-PRB}

\begin{figure}
\includegraphics{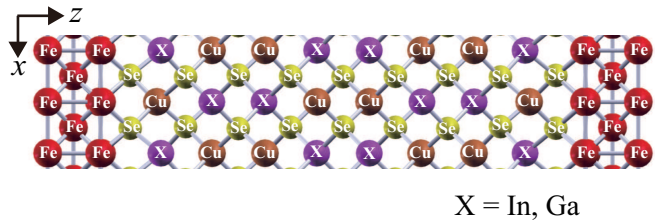}
\caption{\label{supercell}(Color online) Schematic of the supercell used in this study.}
\end{figure}
We prepared supercells of Fe/CIS/Fe and Fe/CGS/Fe (Fig. \ref{supercell}), each of which includes 2 unit cells ($=$17 layers) of CIS or CGS and 1 unit cell ($=$3 layers) of Fe on both sides of the barrier. This barrier length is comparable to $2\,{\rm nm}$ estimated in experiments on the CIGS-based MTJ.\cite{2016Kasai-APL} We fixed the $a$-axis length to $0.5782\, {\rm nm}$ in the case of CIS\cite{1973Parkes-JAC} and $0.5614\, {\rm nm}$ in the case of CGS.\cite{1974Spiess-PSSB} As termination layers, we selected Se layers on the basis of the TEM observation results in the experimental work.\cite{2016Kasai-APL} We next optimized the positions of atoms in the supercells using the density functional theory within the generalized gradient approximation implemented in the Vienna $ab$-$initio$ simulation program (VASP).\cite{1996Kresse-PRB,1999Kresse-PRB} In this optimization, we used a $10\times10\times1$ ${\bf k}$-point mesh and assumed that the spins of all Fe atoms align parallel to each other. As a result of the calculation, the distance between Fe and Se layers in the CIS-based (CGS-based) supercell is determined to be $\sim$$0.167\, {\rm nm}$ ($\sim$$0.144\, {\rm nm}$) in the left boundary and as $\sim$$0.168\, {\rm nm}$ ($\sim$$0.151\, {\rm nm}$) in the right boundary. Such a difference in distance between the left and right boundaries is due to the lack of inversion symmetry along the $c$-axis in CIS and CGS.

To discuss the transport properties of the CIS- and CGS-based MTJs, we consider the quantum open system composed of the above-mentioned supercell attached to the left and right semi-infinite electrodes of Fe atoms. The conductance was calculated with the aid of the quantum code {\rm ESPRESSO}.\cite{Baroni} In the present work, the Coulomb repulsion $U$ for the Cu $3d$ states in the barriers was considered to investigate the change in MR ratio upon changing the amplitude of the band gap systematically.\cite{1991Anisimov-PRB,2013Sclauzero-PRB} First, we obtained the wave functions in each region of the quantum open system by means of the density functional theory and the generalized gradient approximation. The number of ${\bf k}$ points was taken to be $10\times10\times1$, and Methfessel-Paxton smearing with the broadening parameter $0.01\,{\rm Ry}$ was used. The cutoff energies for the wave function and charge density were set to be $30$ and $300\,{\rm Ry}$, respectively. Since our system has a two-dimensional periodicity in the $xy$ plane, the scattering states can be classified by an in-plane wave vector ${\bf k}_{\|} = (k_{x},k_{y})$. For each fixed ${\bf k}_{\|}$ and spin index, we solved the scattering equations derived under the condition that the wave function and its derivative in the supercell are connected to those in the electrodes.\cite{1999Choi-PRB,2004Smogunov-PRB} In this process, we can also obtain the complex band structures, which are useful for understanding the transport properties of the system. Conductance is calculated by substituting the amplitudes of the scattering wave functions into the Landauer formula.

\begin{figure}
\includegraphics{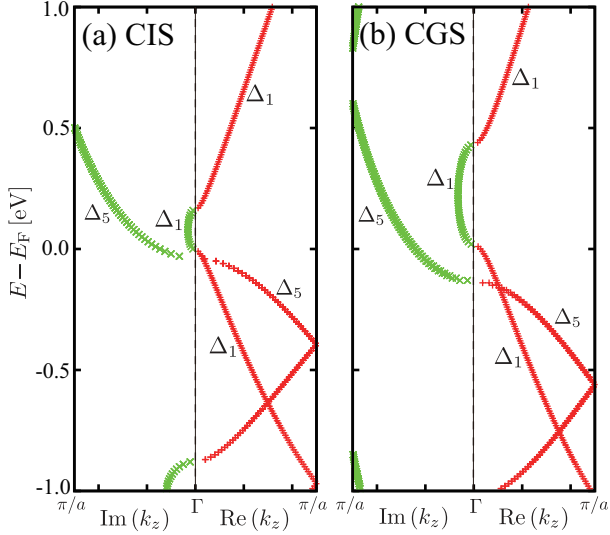}
\caption{\label{complex-band}(Color online) Real and complex band structures of (a) CIS and (b) CGS with $U=5\,{\rm eV}$ at ${\bf k}_{\|}=(0,0)$ along the out-of-plane wave vector $k_z$.}
\end{figure}
We first set the Coulomb repulsion $U$ to $5\,{\rm eV}$ and focused on the difference between the CIS- and CGS-based MTJs. Figures \ref{complex-band}(a) and \ref{complex-band}(b) show the real and complex band structures of the CIS and CGS, respectively, at ${\bf k}_{\|}=(0,0)$ along the out-of-plane wave vector $k_z$. The band gaps are estimated as $E_g^{\rm CIS} \simeq 0.17\,{\rm eV}$ for CIS and $E_g^{\rm CGS} \simeq 0.41\,{\rm eV}$ for CGS. Although these values are smaller than the experimental observations ($E_g^{\rm CIS} \simeq 1.0\,{\rm eV}$ and $E_g^{\rm CGS} \simeq 1.7\,{\rm eV}$), the magnitude relation ($E_g^{\rm CIS} < E_g^{\rm CGS}$) is the same for the present calculations and experiments. We see that the complex band with the $\Delta_{1}$ components has the smallest imaginary part $\kappa_{\rm min} = |{\rm Im} (k_{z})|_{\rm min}$ around the Fermi level in both CIS and CGS. This means that the propagating state with the $\Delta_1$ components in the electrode couples to the evanescent state ($\kappa_{\rm min}$) in the barrier and provides the largest contribution to the tunneling conductance.\cite{2001Butler-PRB}

\begin{figure}
\includegraphics{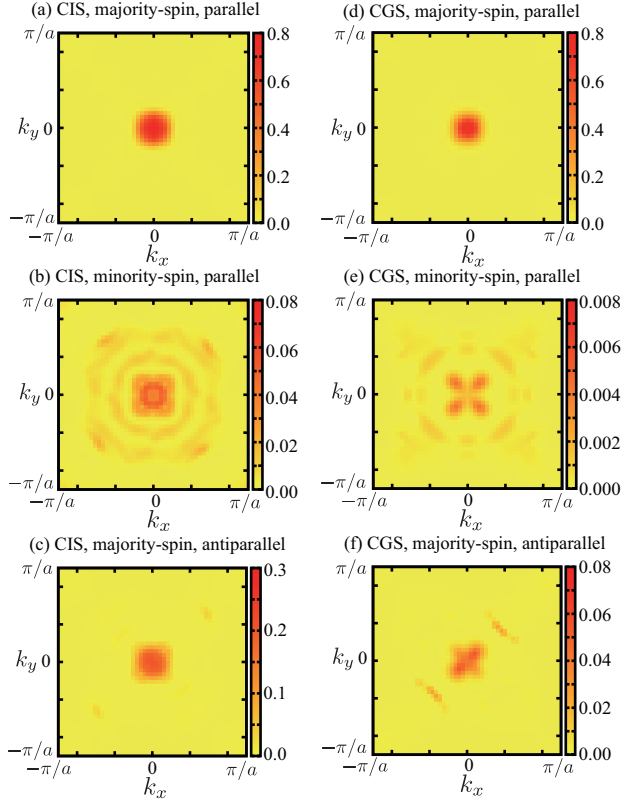}
\caption{\label{k-dep}(Color online) In-plane wave vector ${\bf k}_{\|}=(k_x,k_y)$ dependence of the conductances at the Fermi energy for various cases in CIS- and CGS-based MTJs with $U=5\,{\rm eV}$. (See the text for details.) The unit of the color bar is $G_0=e^2/h$ in all panels.}
\end{figure}
In Fig. \ref{k-dep}, we show the in-plane wave vector ${\bf k}_{\|}=(k_x,k_y)$ dependence of the conductances at the Fermi energy for various situations in the CIS- and CGS-based MTJs. The upper two panels, Figs. \ref{k-dep}(a) and \ref{k-dep}(d), show the majority-spin conductances of the CIS- and CGS-based MTJs with the parallel magnetization of the electrodes, in which the sharp peaks around ${\bf k}_{\|}=(0,0)$ are seen for both MTJs. As shown in Figs. \ref{complex-band}(a) and \ref{complex-band}(b), since the $\Delta_1$ evanescent state is the dominant conducting channel at ${\bf k}_{\|}=(0,0)$ in the barriers, these peaks can be considered as evidence of the tunneling transport by the $\Delta_1$ wave functions. Figures \ref{k-dep}(b) and \ref{k-dep}(e) show the minority-spin conductances of the CIS- and CGS-based MTJs with the parallel magnetization of the electrodes. Compared with the majority-spin cases, conductances have much smaller values and are distributed over a wide region of the ${\bf k}_{\|}$ Brillouin zone for both CIS- and CGS-based MTJs. The lower two panels, Figs. \ref{k-dep}(c) and \ref{k-dep}(f), show the conductances of the CIS- and CGS-based MTJs for the majority-spin states of the left electrodes in the case of the antiparallel magnetization. Although the $\Delta_1$ wave function in the left electrode decays more rapidly than in the case of parallel magnetization, it still has a small amplitude in the right electrode after passing through the barrier, which is the origin of the small conductances around ${\bf k}_{\|}=(0,0)$. Note that the ${\bf k}_{\|}$ dependences in Figs. \ref{k-dep}(c) and \ref{k-dep}(f) break the fourfold rotational symmetry, which might be due to the twofold symmetry of CIS and CGS in the $xy$ plane. Although not shown here, the minority-spin conductances in the case of antiparallel magnetization have ${\bf k}_{\|}$ dependences that are identical to those obtained by rotating conductance distributions in Figs. \ref{k-dep}(c) and \ref{k-dep}(f) by $180^\circ$ in the ${\bf k}_{\|}$ plane.

In this work, we adopt the usual optimistic definition of the MR ratio: ${\rm MR\, ratio\, [\%]} = 100 \times (T_{\rm P}-T_{\rm AP})/T_{\rm AP}$, where $T_{\rm P}$ ($T_{\rm AP}$) is the sum of the majority- and minority-spin conductances averaged over the ${\bf k}_{\|}$ Brillouin zone in the case of parallel (antiparallel) magnetization. We obtained 62.3 and $300.9\%$ MR ratios for the CIS- and CGS-based MTJs, respectively. The difference is mainly due to a large difference in conductance in the case of antiparallel magnetization [Figs. \ref{k-dep}(c) and \ref{k-dep}(f)]. We also estimated $RA$ values from the conductances in the case of the parallel magnetization $T_{\rm P}$, where 0.408 and $0.680\, \Omega\, {\mu {\rm m}}^2$ were obtained for the CIS- and CGS-based MTJs, respectively. In the present study, the in-plane lattice constant of Fe/CIS(CGS)/Fe supercells is twice as long as that of bcc Fe (see Fig. \ref{supercell}). Since CIS and CGS have quite small but finite displacements in Se atoms from fractional positions,\cite{1984Jaffe-PRB} we need to use a larger supercell as compared with other semiconductors (Si, GaAs or ZnSe). In such a larger supercell, the band folding of Fe can reduce MR ratios, as discussed for Fe/MgAl$_2$O$_4$/Fe(001) MTJs.\cite{2012Miura-PRB,2012Sukegawa-PRB} However, in the present case, the folded band hardly affects MR ratios of the CIS- and CGS-based MTJs. Actually, we confirmed that the folded minority-spin band of Fe crossing the Fermi level provides no contribution to the conductance in both MTJs because of the very small displacements of Se atoms.

\begin{figure}
\includegraphics{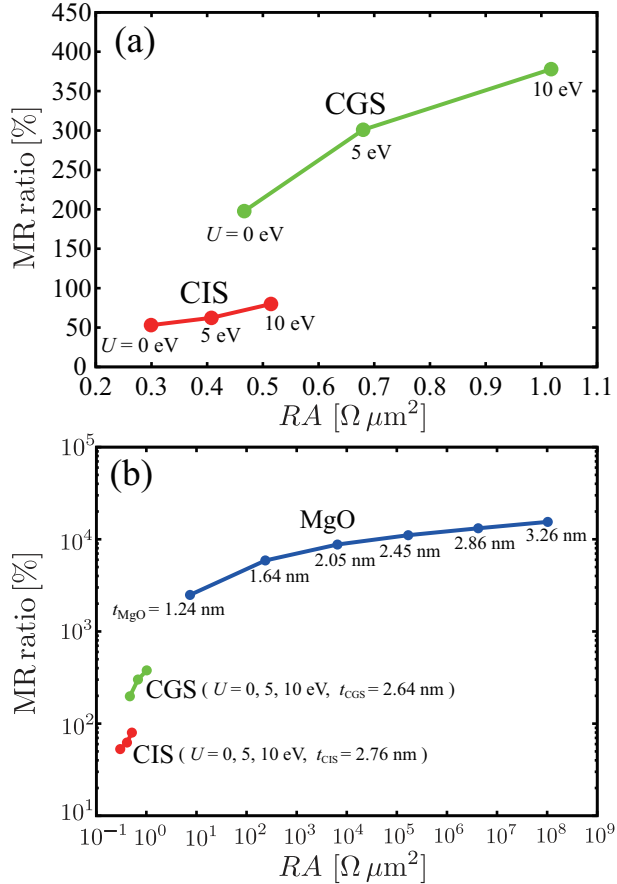}
\caption{\label{RAvsMR}(Color online) MR ratios and $RA$ values of (a) CIS- and CGS-based MTJs and (b) CIS-, CGS-, and MgO-based MTJs on a double-logarithmic scale. In the panel (b), barrier thickness ($t_{\rm CIS}$, $t_{\rm CGS}$, or $t_{\rm MgO}$) is defined as the distance between two Fe layers closest to the barrier.}
\end{figure}
Let us further discuss the relationship between the band gaps and MR ratios by changing the Coulomb repulsion $U$ for the Cu $3d$ states in the barriers. By increasing the repulsion $U$ from $0$ to $10\,{\rm eV}$, the band gap of bulk CIS (CGS) is increased from $0.044$ ($0.114$) to $0.468$ ($0.691$) eV. In Fig. \ref{RAvsMR}(a), we show the MR ratios and $RA$ values of the CIS- and CGS-based MTJs with $U=0,5,$ and $10\,{\rm eV}$. As the repulsion $U$ becomes larger, the MR ratio and $RA$ increase in both MTJs. It is also found that for a fixed repulsion $U$, the CGS-based MTJ has a higher MR ratio and a larger $RA$ than the CIS-based MTJ. From these, we can conclude, at least for the CIS- and CGS-based MTJs, that a larger gap system has a higher MR ratio and a larger $RA$.

Figure \ref{RAvsMR}(b) shows the MR ratios and $RA$ values of the CIS-, CGS-, and MgO-based MTJs. The MR ratio and $RA$ of the MgO-based MTJ are calculated in the same manner as those of the CIS- and CGS-based MTJs. For the barrier thickness of 2.6$-$2.8$\, {\rm nm}$, the $RA$ values of the CIS- and CGS-based MTJs are nearly six orders of magnitude smaller than those of the MgO-based MTJs, which is a great advantage of the CIS- and CGS-based MTJs for certain device applications where a small $RA$ is needed, e.g., read sensors. Even if we reduce the thickness of the MgO to 1.24 nm, $RA$ is still larger than those of the CIS and CGS systems. On the other hand, the CIS- and CGS-based MTJs have the possibility of achieving even smaller $RA$ values by reducing their barrier thicknesses. In that case, establishing a method to keep the MR ratio high, e.g., the use of highly spin-polarized ferromagnetic electrodes, is essential. In fact, such an example has recently been reported by Kasai {\it et al.}\cite{2016Kasai-APL}

Finally, let us discuss the reason why only low MR ratios have been observed in the GaAs- and ZnSe-based MTJs,\cite{2002Kreuzer-APL,2006Moser-APL,2003Jiang-APL} as opposed to the CIGS-based MTJ. To this end, we calculated MR ratios of Fe/GaAs/Fe and Fe/ZnSe/Fe MTJs using supercells with 17 barrier layers and 3 Fe layers on both sides of the barrier. The termination layers of GaAs and ZnSe were determined to be As and Se layers, respectively, from the energy minimization of the supercells. All conditions in the transport calculations were the same as those for the CIS- and CGS-based MTJs. In the GaAs-based MTJ, we obtained an MR ratio of $12\%$, which is more than one order smaller than the value reported in the previous theoretical work.\cite{2010Autes-PRB} Such a difference comes from the sensitivity of the MR ratio in the GaAs-based MTJ to the position of the Fermi level.\cite{2010Autes-PRB} In general, the Fermi level of magnetic junctions strongly depends on the in-plane lattice constant, the interfacial distance, and other calculation conditions, leading to different positions of the Fermi level in each calculation. In our case, the Fermi level is located near the interfacial resonance states of Fe, which yields a low MR ratio. On the other hand, in Ref. \citen{2010Autes-PRB}, the Fermi level was set to the middle of the band gap in the barrier, where the effect of the interfacial resonance states is small and a high MR ratio is obtained. As mentioned in the introduction, experiments on Fe/GaAs/Fe MTJs have revealed low MR ratios ($<$$2\%$),\cite{2002Kreuzer-APL,2006Moser-APL} which can be explained as an effect of the interfacial resonance states. In the case of the ZnSe-based MTJ, we obtained a high MR ratio of $292\%$, which is consistent with the previous theoretical estimation.\cite{1999MacLaren-PRB} We also found that the interfacial resonance states do not provide a significant contribution to the MR ratio in this system. Therefore, a low MR ratio ($\sim$$10\%$) observed in experiments\cite{2003Jiang-APL} would be due to experimental imperfections. Actually, the existence of a large film roughness ($\sim$$0.9\,{\rm nm}$) is confirmed in Ref. \citen{2003Jiang-APL}.

In summary, we studied transport properties of MTJs with semiconductor barriers, Fe/CIS/Fe and Fe/CGS/Fe. Our first-principles-based calculations showed that $\Delta_1$ wave functions dominate the tunneling transport in both MTJs. The theoretical transport calculations predicted MR ratios of around $50\%$ for the CIS-based MTJ and around $300\%$ for the CGS-based one, which means that a higher MR ratio is expected for a higher Ga concentration $x$ in ${\rm CuIn}_{1-x}{\rm Ga}_{x}{\rm Se}_{2}$-based MTJs. We further discussed the relationship between the band gap and MR ratio by changing the Coulomb repulsion in the CIS and CGS barriers. We found that a larger band gap in the barrier gives a higher MR ratio. Through the comparison with the MgO-based MTJ, we confirmed that the CIS- and CGS-based MTJs have much smaller $RA$ values than the MgO-based MTJ, which is consistent with the experimental results of the ${\rm CuIn}_{0.8}{\rm Ga}_{0.2}{\rm Se}_{2}$-based MTJ.\cite{2016Kasai-APL}

\begin{acknowledgment}
The authors are grateful to K. Hono, S. Kasai, and K. Mukaiyama for useful discussions and critical comments. This work was in part supported by Grant-in-Aids for Scientific Research (S) (Grant No. 16H06332) and (B) (Grant No. 16H03852) from the Ministry of Education, Culture, Sports, Science and Technology, Japan, by NIMS MI2I, and also by the ImPACT Program of Council for Science, Technology and Innovation, Japan.
\end{acknowledgment}

\end{document}